\documentclass[11pt,showpacs,preprintnumbers,superscriptaddress,amsmath,amssymb,nofootinbib]{revtex4}

\usepackage{graphicx}% Include figure files
\usepackage{dcolumn}% Align table columns on decimal point
\usepackage{bm}% bold math
\usepackage{amssymb}
\usepackage{amsmath}
\usepackage{epsfig}    
\usepackage{color}
\usepackage{slashed}
\def\be{\begin{equation}}
\def\ee{\end{equation}}
\newcommand{\bea}{\begin{eqnarray}}
\newcommand{\eea}{\end{eqnarray}}

\numberwithin{equation}{section}

\begin{document} 

%\title{Phenomenology of Two Higgs Doublet Models with the $S_3$ Flavor Symmetry and Its Application to Hybrid Seesaw Model}
\title{Electron/Muon Specific Two Higgs Doublet Model}
\preprint{KIAS-P13053}

\author{Yuji Kajiyama}
\email{kajiyama-yuuji@akita-pref.ed.jp}
\affiliation{Akita Highschool, Tegata-Nakadai 1, Akita, 010-0851, Japan}
\author{Hiroshi Okada}
\email{hokada@kias.re.kr}
\affiliation{School of Physics, KIAS, Seoul 130-722, Korea}
\author{Kei Yagyu}
\email{keiyagyu@ncu.edu.tw}
\affiliation{Department of Physics, National Central University, Chungli, Taiwan 32001, ROC}

\begin{abstract}

We discuss two Higgs doublet models with a softly-broken discrete $\mathbb{S}_3$  symmery, where  
the mass matrix for charged-leptons is predicted as the diagonal form in the weak eigenbasis of lepton fields.
Similar to an introduction of $\mathbb{Z}_2$ symmetry, 
the tree level flavor changing neutral current can be forbidden by imposing the $\mathbb{S}_3$ symmetry to the model.
Under the $\mathbb{S}_3$ symmetry,  
there are four types of Yukawa interactions depending on the $\mathbb{S}_3$ charge assignment to right-handed fermions. 
We find that extra Higgs bosons can be muon and electron specific in one of four types of the Yukawa interaction. 
This property does not appear in any other two Higgs doublet models with a softly-broken ${\mathbb Z}_2$ symmetry. 
We discuss the phenomenology of the muon and electron specific Higgs bosons at the Large Hadron Collider; namely 
we evaluate allowed parameter regions from the current Higgs boson search data and 
discovery potential of such a Higgs boson at the 14 TeV run. 

\end{abstract}
\maketitle
\newpage

\section{Introduction}

A Higgs boson has been discovered at the CERN Large Hadron Collider (LHC)~\cite{ATLAS_Higgs, CMS_Higgs}, whose properties; $e.g.$, mass, spin, CP and observed number of events 
are consistent with those of the Higgs boson predicted in the Standard Model (SM). 
The SM-like Higgs boson also appears in Higgs sectors extended from the SM one, so that 
there are still various possibilities non-minimal Higgs sectors. 
They are often introduced in models beyond the SM which have been considered 
to explain problems unsolved within the SM such as the neutrino oscillation, dark matter (DM) and baryon asymmetry of the Universe.  

In addition to the above problems, one of the deepest mystery in the SM is the flavor structure. 
In the SM, all the masses of charged fermions are accommodated 
by the vacuum expectation value (VEV) of the Higgs doublet field through Yukawa interactions. 
However, there are redundant number of parameters to obtain physical observables; $i.e.,$
the Yukawa couplings are given by general $3\times 3$ complex matrices (totally 18 degrees of freedom)
for each up-type and down-type quarks and charged-leptons. 
In fact, only three independent parameters are enough in the charged-leptons sector to describe the masses of $e$, $\mu$ and $\tau$. 
In order to constrain the structure of Yukawa interactions, 
Non-Abelian discrete symmetries have been 
introduced such as based on the $\mathbb{S}_3$~\cite{S3,S3_2} and $A_4$~\cite{A4} groups. 
Usually, in a model with such a discrete symmetry, 
the Higgs sector is extended to be the multi-doublet structure.
Therefore, phenomenological studies for the extended Higgs sector with multi-doublet structure are important to probe such a model.

In this paper, we discuss two Higgs doublet models (THDMs) with the $\mathbb{S}_3$ symmetry as the simplest realization of 
the diagonalized mass matrix for the charged-leptons without introducing any unitary matrices. 
This can be achieved by assigning the first and second generation lepton fields 
to be the $\mathbb{S}_3$ doublet\footnote{
Our $\mathbb{S}_3$ charge assignments for the quarks and Higgs doublet fields are 
different from those in the previous studies for $\mathbb{S}_3$ models~\cite{S3_2}. 
Usually, all the quarks, leptons and Higgs doublet fields are embedded in the $\mathbb{S}_3$ doublet plus singlet. 
However, we treat that the quark sector is the same as in the SM assuming the quark fields to be the singlet, because  
it is suitable and economical to explain the observed SM-like Higgs boson at the LHC. }.  

In general, there appears the flavor changing neutral current (FCNC) via a neutral Higgs boson mediation at the tree level 
in two Higgs doublet models (THDMs), which is strictly constrained by flavor experiments. 
Usually, such a tree level FCNC is forbidden by introducing a discrete $\mathbb{Z}_2$ symmetry~\cite{Glashow} to realize the 
situation where one of two Higgs doublet fields couples to each fermion. 
In our model, this situation is realized in terms of the $\mathbb{S}_3$ flavor symmetry. 
The Yukawa interaction among the Higgs doublet fields and fermions can be classified into four types 
depending on the $\mathbb{S}_3$ charge assignments to the right-handed fermions. 
Similar classification has been defined in THDMs with a softly-broken $\mathbb{Z}_2$ symmetry~\cite{z2,Akeroyd}. 
A comprehensive review for the THDMs with the softly-broken $\mathbb{Z}_2$ symmetry has been given in Ref.~\cite{THDM_Rev}.

We find that extra neutral and charged Higgs bosons can be muon and electron specific; namely,  
they can mainly decay into $\mu^+\mu^-$ or $e^+e^-$ and $\mu^\pm\nu$ or $e^\pm\nu$, respectively, 
in one of four types of the Yukawa interaction. 
This phenomena cannot be seen in any other THDMs without the tree level FCNC such as the 
softly-broken $\mathbb{Z}_2$ symmetric version. 
We show excluded parameter regions from the current LHC data in this scenario. 
We then evaluate discovery potential of signal events from these extra Higgs bosons at the LHC with the collision energy to be 14 TeV.

%[Organization]

This paper is organized as follows.
In Sec.~II, we define the particle content and give the Lagrangian in our model. 
The mass matrices for the charged-leptons and neutrinos are then calculated. 
The Higgs boson interactions are also discussed in this section. 
In Sec.~III, we discuss the collider phenomenology, especially focusing on the 
muon and electron specific Higgs bosons in the Type-I $\mathbb{S}_3$ model. 
We give a summary and conclusion of this paper in Sec.~IV.
The explicit form of the Higgs potential and mass matrices for the scalar fields are given in Appendix~A.

%%%%%%%%%%%%%%%%%%%%%%%%%%%%%%%%%%%%%%
\section{The Model}

\subsection{Charge assignment}
\begin{table}[t]
\centering {\fontsize{10}{12}
\begin{tabular}{c|c|c|c|c|c|c|c|c|c}
\hline\hline Particle &$Q_i$& $L_{a}$ & $L_\tau$ &$u_{iR}$&$d_{iR}$& $ e_{aR} $ & $ \tau_R $  
& $\Phi_1 $& $\Phi_2 $  \\\hline
$SU(2)_L,U(1)_Y$& $\bm{2},1/6$ & $\bm{2},-1/2$ & $\bm{2},-1/2$ &$\bm{1},2/3$&$1,-1/3$& $\bm{1},-1$& $\bm{1},-1$ &  $\bm{2},1/2$
& $\bm{2},1/2$    \\\hline
$\mathbb{S}_3$ & $\mathbf{1}$ & $\mathbf{2}$ & $\mathbf{1}$ & $\mathbf{1'}$ &$\mathbf{1}$ or $\mathbf{1'}$& $\mathbf{2}$& $\mathbf{1}$ or $\mathbf{1'}$ & $\mathbf{1}$&  $\mathbf{1'}$ \\
\hline\hline
\end{tabular}%
} 
\caption{The particle contents and their charge assignment of the $SU(2)_L\times U(1)_Y\times \mathbb{S}_3$ symmetry. }
\label{tab:1}
\end{table}

\begin{table}[t]
\centering {\fontsize{10}{12}
\begin{tabular}{c|c|c|c||c|c|c}
\hline\hline ~~Particle~~ &  ~~$u_{iR}$~~ &~~$d_{iR}$~~ &~~$\tau_R$~~ & ~~$\xi_u$~~ & ~~$\xi_d$~~ & ~~$\xi_\tau$~~  \\\hline
Type-I &  $1'$&$1'$&$1'$ &$\cot\beta$&$\cot\beta$&$\cot\beta$  \\\hline
Type-II & $1'$&$1$&$1$  &$\cot\beta$&$-\tan\beta$&$-\tan\beta$\\\hline
Type-X &  $1'$&$1'$&$1$ &$\cot\beta$&$\cot\beta$&$-\tan\beta$\\\hline
Type-Y &  $1'$&$1$&$1'$ &$\cot\beta$&$-\tan\beta$&$\cot\beta$\\\hline\hline
\end{tabular}%
} 
\caption{Four patterns of the assignment of $\mathbb{S}_3$ charges to the right-handed fermions, and $\xi_f$ factors appearing in Eq.~(\ref{int}).}
\label{tab:2}
\end{table}

We discuss the THDM with the softly-broken discrete $\mathbb{S}_3$ symmetry. 
In the $\mathbb{S}_3$ group, there are the following irreducible representations; 
two singlets $\mathbf{1}$ (true-singlet) and $\mathbf{1'}$ (pseudo-singlet) and doublet $\mathbf{2}$ (see Ref.~\cite{Ma:2013zca}). 
Particle contents are shown in Table~\ref{tab:1}. 
The $i$-th generation of left-handed quarks $Q_i$ are assigned to be $\mathbb{S}_3$ true-singlet, while
the right-handed up type quarks $u_{iR}$ and down type quarks $d_{iR}$ are assigned to be $\mathbb{S}_3$ true- or pseudo-singlet. 
The left- (right-) handed electron and muon $L_a$ ($e_{aR}$) 
are embedded as the doublet representation of the $\mathbb{S}_3$ symmetry.   
Both left-handed and right-handed tau leptons $L_\tau$ and $\tau_R$, respectively, are singlets under $\mathbb{S}_3$. 
The isospin doublet Higgs fields $\Phi_1$ and $\Phi_2$ are transformed as $\mathbb{S}_3$ true- or pseudo-singlet. 

We can define four independent patterns of the charge assignment for $u_{iR}$, $d_{iR}$ and $\tau_{R}$ 
in the $\mathbb{S}_3$ symmetric THDMs.  
We call them as Type-I, Type-II, Type-X and Type-Y $\mathbb{S}_3$ models, and the $\mathbb{S}_3$ charge assignment in each model is listed in Table~\ref{tab:2}.  
This charge assignment\footnote{
The Type-X and Type-Y THDMs are respectively referred as the lepton-specific~\cite{lepton_specific} and flipped~\cite{flipped} THDMs.} is the analogy of that of a softly-broken $\mathbb{Z}_2$ symmetry in the THDMs~\cite{typeX}.

\subsection{Higgs Potential}

The softly-broken $\mathbb{S}_3$ symmetric Higgs potential is given as
\begin{align}
V&=m_1^2 \Phi^\dag_1 \Phi_1 + m_2^2 \Phi^\dag_2 \Phi_2 + \left[m_3^2\Phi^\dag_1 \Phi_2 + \text{h.c.}  \right]\notag\\
&+\frac{1}{2}\lambda_1(\Phi^\dag_1 \Phi_1)^2
+\frac{1}{2}\lambda_2(\Phi^\dag_2 \Phi_2)^2
+\lambda_3(\Phi_1^\dagger \Phi_1)(\Phi_2^\dagger \Phi_2)
+\lambda_4|\Phi^\dag_1 \Phi_2|^2
+\frac{1}{2}\left[\lambda_5(\Phi^\dag_1 \Phi_2)^2+{\rm h.c.}\right], 
\label{pot_gen}
\end{align}
where 
the doublet Higgs fields can be parameterized as 
\begin{align}
\Phi_\alpha &=\left[
\begin{array}{c}
w_\alpha^+\\
\frac{1}{\sqrt{2}}(h_\alpha+v_\alpha+iz_\alpha)
\end{array}\right],~ \alpha = 1,2,
\end{align}
where $v_\alpha$ are the VEVs of the doublet Higgs fields, and they satisfy $v^2\equiv v_1^2+v_2^2=
1/(\sqrt{2}G_F)=(246$ GeV$)^2$. 
The ratio of the two VEVs can be parameterized by $\tan\beta\equiv v_2/v_1$ as usual in THDMs.  
Although among the parameters in the potential, $m_3^2$ and $\lambda_5$ are complex in general, 
we assume the CP-conservation in the Higgs potential for simplicity. 
We note that 
we can retain the $\mathbb{Z}_2$ symmetry as the subgroup of $\mathbb{S}_3$ by taking $m_3^2=0$. 
However, the potential without the $m_3^2$ term results non-decoupling theory; namely,  
all the masses of Higgs bosons are determined by the Higgs VEV times $\lambda$ couplings. 
In the following, we consider the case with $m_3^2\neq 0$. 

The mass eigenstates for the CP-odd, singly-charged and CP-even Higgs bosons from the doublet fields 
are given by the following orthogonal matrices as
\begin{align}
\left(
\begin{array}{c}
z_1\\
z_2
\end{array}\right)&=
\left(\begin{array}{cc}
c_\beta & -s_\beta  \\
s_\beta & c_\beta  
\end{array}\right)
\left(
\begin{array}{c}
G^0\\
A\\
\end{array}\right),~
\left(
\begin{array}{c}
w_1^+\\
w_2^+
\end{array}\right)=
\left(\begin{array}{cc}
c_\beta & -s_\beta \\
s_\beta & c_\beta 
\end{array}\right)
\left(
\begin{array}{c}
G^+\\
H^+
\end{array}\right),\notag\\
\left(
\begin{array}{c}
h_1\\
h_2
\end{array}\right)&=
\left(\begin{array}{cc}
c_\alpha & -s_\alpha \\
s_\alpha & c_\alpha
\end{array}\right)
\left(
\begin{array}{c}
H\\
h
\end{array}\right), 
\end{align}
where $G^\pm$ and $G^0$ are the Nambu-Goldstone bosons
which are absorbed by the longitudinal component of $W^\pm$ and $Z$. 
Because the potential given in Eq.~(\ref{pot_gen}) is the completely same form as in the softly-broken $\mathbb{Z}_2$ symmetric 
THDMs, the mass formulae are also the same form.  
The detailed formulae for the masses of the physical Higgs bosons can be seen in Ref.~\cite{THDM_Chiang-Yagyu}, for example.

\subsection{Yukawa Lagrangian}

The renormalizable Yukawa Lagrangian under the $\mathbb{S}_3$ invariance is given by
\begin{align}
-\mathcal{L}_{Y}
&=
y_1^{\ell}(\bar{L}_1e_{2R}+\bar{L}_2 e_{1R} )\Phi_1  + y_2^{\ell} (\bar{L}_1 e_{2R} -\bar{L}_2 e_{1R} )\Phi_2
+\text{h.c.}\notag\\
&+y^u_{ij}\bar{Q}_i(i\tau_2\Phi_u^*) u_{jR}+y^d_{ij}\bar{Q}_i\Phi_d d_{jR}+y^\tau\bar{L}_\tau\Phi_\tau \tau_R+\text{h.c.}, 
\label{main-lag}
\end{align}
where $\Phi_{u,d,\tau}$ are $\Phi_{1}$ or $\Phi_2$ depending on the $\mathbb{S}_3$ charge assignment of 
$u_{iR}^{}$, $d_{iR}^{}$ and $\tau_R^{}$ as listed in Table~\ref{tab:2}.

The charged-lepton mass matrix defined by $(\bar{e}_L,\bar{\mu}_L,\bar{\tau}_L)M_\ell (e_R,\mu_R,\tau_R)^T$, under the 
identifications of the lepton fields as
$L_1=L_e,~L_2=L_\mu,~e_{1R}=\mu_{R},~e_{2R}=e_R$, 
can be obtained in the diagonal form by 
\begin{align}
M_\ell =
\frac{1}{\sqrt{2}}\text{diag}(y_1^\ell v_1+y_2^\ell v_2,~y_1^\ell v_1-y_2^\ell v_2,~y^\tau v_\tau), 
\end{align} 
where $v_\tau$ is either $v_1$ or $v_2$. 

The quarks masses and mixings are obtained as the same way in the SM. 
As already mentioned in the Introduction, this treatment is different from that in the previous $\mathbb{S}_3$ models~\cite{S3_2} in which the part of Yukawa Lagrangian is 
given by the $\mathbb{S}_3$ singlet from $\mathbf{2}\times\mathbf{2}\times \mathbf{2}$, where each $\mathbf{2}$ denotes the left-handed quark, right-handed quark and Higgs doublet fields. 
In such a model, there are predictions in the quark sector such as the Cabibbo mixing angle. 
In our model, we choose singlet representations for all the quark fields and Higgs doublet fields, so that 
there is no such a prediction. 
However, by this assignment, the minimal content for the Higgs sector; $i.e.,$ two Higgs doublet fields 
can be realized within the framework of $\mathbb{S}_3$ with the diagonalized 
charged-lepton mass matrix and the SM-like Higgs boson which is necessary to explain the observed Higgs boson at the LHC 
as will be discussed in the next subsection.

The Yukawa interactions are given in the mass eigenbasis for the physical Higgs bosons as  
\begin{align}
-{\mathcal L}_Y^\text{int}
&=\frac{m_\mu}{v}\Big\{-\frac{1}{2}(\tan\beta+\cot\beta)c_{\beta-\alpha}\bar{e}eh 
+\Big[s_{\beta-\alpha}-\frac{1}{2}(\tan\beta-\cot\beta)c_{\beta-\alpha}\Big]\bar{\mu}\mu h \notag\\
&\hspace{1.5cm}+\frac{1}{2}(\tan\beta+\cot\beta)s_{\beta-\alpha}\bar{e}eH
+\Big[c_{\beta-\alpha}+\frac{1}{2}(\tan\beta-\cot\beta)s_{\beta-\alpha}\Big]\bar{\mu}\mu H \notag\\
&\hspace{1.5cm}-\frac{i}{2}(\tan\beta+\cot\beta)\bar{e}\gamma_5e A
-\frac{i}{2}(\tan\beta-\cot\beta)\bar{\mu}\gamma_5\mu A\notag\\
&\hspace{1.5cm}-\frac{1}{\sqrt{2}}\Big[ (\tan\beta+\cot\beta)\bar{\nu}_eP_ReH^++
(\tan\beta-\cot\beta)\bar{\nu}_\mu P_R \mu H^++\text{h.c.}\Big]\Big\}\notag\\
&\sum_{f=u,d,\tau}\frac{m_f}{v}\Big[ (s_{\beta-\alpha}+\xi_f c_{\beta-\alpha}){\overline
f}fh+(c_{\beta-\alpha}-\xi_f s_{\beta-\alpha}){\overline f}fH-2iI_f \xi_f{\overline f}\gamma_5fA\Big]\notag\\
&\hspace{1.5cm}+\Big[\frac{\sqrt2V_{ud}}{v}\overline{u}
\left( m_d\xi_d P_R - m_u\xi_u P_L \right)d\,H^+
+\frac{\sqrt2m_\tau\xi_\tau}{v}\overline{\nu_\tau^{}}P_R\tau^{}H^+
+\text{h.c.}\Big], \label{int}
\end{align}
where the electron mass is neglected in the above expression,
and $I_f=+1/2~(-1/2)$ for $f=u$ ($d,\tau$).  
The $\xi_f$ factors are listed in Table~\ref{tab:2}.  

The $hVV$ and $HVV$ ($V=W^\pm,Z$) coupling constants are 
given by $\sin(\beta-\alpha)\times g_{hVV}^{\text{SM}}$ and $\cos(\beta-\alpha)\times g_{hVV}^{\text{SM}}$ 
with $g_{hVV}^{\text{SM}}$ being the coupling constant of the SM Higgs boson and gauge bosons. 
Thus, when we take the limit of $\sin(\beta-\alpha)=1$, $h$ has the same coupling constants with 
the gauge bosons and fermions (see Eq.~(\ref{int})) as those in the SM Higgs boson.  

We here comment on the new contributions to the muon anomalous magnetic moment ($g-2$) from the additional 
scalar boson loops. 
In our model in the case of $\sin(\beta-\alpha)=1$, 
the $H$, $A$ and $H^\pm$ loop contributions are calculated by using the formula given in Ref.~\cite{g-2_strumia} as
\begin{align}
\Delta a_\mu &= \frac{1}{32\pi^2}\frac{m_\mu^2}{v^2}(\tan\beta-\cot\beta)^2\notag\\
&\Big\{
[F_1(m_H^2/m_\mu^2)+F_2(m_H^2/m_\mu^2)]+[-F_1(m_A^2/m_\mu^2)+F_2(m_A^2/m_\mu^2)]-\frac{m_\mu^2}{6m_{H^+}^2}\Big\}, 
\end{align}
where 
\begin{align}
F_1(x) &=\frac{1-4x+3x^2-2x^2\ln x}{2(1-x)^3},\quad
F_2(x)=-\frac{(1-x)(2x^2+5x-1)+6x^2\ln x}{6(1-x)^4}. 
\end{align}
The numerical values derived from the above formula agree with those using formula given in Ref.~\cite{g-2_2hdm}. 
When we only take into account the $H$ loop contribution, and we set $m_H=150$ (300) GeV, the numerical value is obtained about
$3\times 10^{-11}~(9\times 10^{-12})~\times \tan^2\beta/100$. The $A$ and $H^\pm$ loops give destructive contributions to the 
$H$ loop contribution. 
On the other hand, the discrepancy of the measured muon $g-2$ from the SM prediction is roughly 
given as $3\times 10^{-9}$~\cite{discrepancy1,discrepancy2} which is two orders of magnitude larger than the above result with 
$m_H=150$ GeV and $\tan\beta=100$. 
Therefore, it is difficult to compensate the discrepancy by the additional scalar boson loop contributions in our model similar to the 
Type-II THDM. 

\section{Phenomenology at the LHC}

In this section, we discuss the phenomenology of the Higgs bosons at the LHC. 
We consider the case with $\sin(\beta-\alpha)=1$ in which $h$ can be regarded as the SM-like Higgs boson with 
the mass of 126 GeV, because the current Higgs boson search data at the LHC suggest that the observed Higgs boson 
is consistent with the SM Higgs boson. 
We then focus on collider signatures from the extra Higgs bosons; $i.e.$, $H$, $A$ and $H^\pm$ at the LHC.

\subsection{The $\mu$ and $e$ specific Higgs bosons}

In all the $\mathbb{S}_3$ models defined in Table~\ref{tab:2}, the coupling constants of the extra Higgs bosons with $\mu$ and $e$ are respectively proportional to 
($\tan\beta-\cot\beta$) and ($\tan\beta+\cot\beta$) as seen in Eq.~(\ref{int}). 
Thus, the extra Higgs bosons are expected to be $\mu$ and $e$ specific in large or small $\tan\beta$ regions\footnote{
Cases with small $\tan\beta$; $i.e.,$ $\tan\beta\lesssim 1$ is typically disfavored by the $B$ physics data such as the $b\to s\gamma$ process~\cite{Misiak,Stal,Haisch}. }. 
However, this feature is hidden in the Type-II, Type-X and Type-Y $\mathbb{S}_3$ models, because at least one of the bottom or tau Yukawa couplings
is also enhanced as getting larger values of $\tan\beta$. 
Therefore, 
phenomenology in the Type-II, Type-X and Type-Y $\mathbb{S}_3$ models are almost the same as those in the 
Type-II, Type-X and Type-Y THDMs with the softly-broken $\mathbb{Z}_2$ symmetry, respectively. 
Studies for collider signatures using data of 126 GeV Higgs boson at the LHC have been analyzed in Refs.~\cite{THDM_Chiang-Yagyu,THDM_Higgs} in the softly-broken $\mathbb{Z}_2$ symmetric THDMs.
Only in the Type-I $\mathbb{S}_3$ model, 
all the Yukawa couplings of the extra Higgs bosons are suppressed by $\cot\beta$, 
so that the $\mu$ and $e$ specific nature is maintained. 

We would like to emphasize that 
appearance of the $\mu$ and $e$ specific extra Higgs bosons 
does not appear in the other THDMs without the tree level FCNC; $e.g.$,
the $\mathbb{Z}_2$ symmetric version and the THDMs with Yukawa alignments discussed in Ref.~\cite{pich}.  
In such a THDM, 
the interaction matrices among a Higgs boson and fermions are proportional to the fermion mass matrices. 
Therefore, the branching fractions of $\mathcal{H}\to\mu\mu$ and $\mathcal{H}\to ee$ are suppressed by the factors of 
$(m_\mu/m_\tau)^2$ and $(m_e/m_\tau)^2$, respectively, compared to that of $\mathcal{H}\to\tau\tau$, 
where $\mathcal{H}$ denotes an extra neutral Higgs boson.
If we consider the most general THDM, sometimes it is called as the Type-III THDM~\cite{typeIII},  
in which both Higgs doublet fields couple to each fermion, such a proportionality between the matrices 
can be broken in general. 
In that case, the $\mu$ and $e$ specific extra Higgs bosons can be obtained by choosing parameters in the interaction matrix. 
The important point in our model is that we can explain the $\mu$ and $e$ specific nature
as a consequence of the $\mathbb{S}_3$ symmetry. 

Therefore, measuring signatures from the $\mu$ and $e$ specific extra Higgs bosons can be useful to distinguish the other 
THDMs without the tree level FCNC. 

\subsection{Decays of extra Higgs bosons}

\begin{figure}[t]
\begin{center}
 \includegraphics[width=70mm]{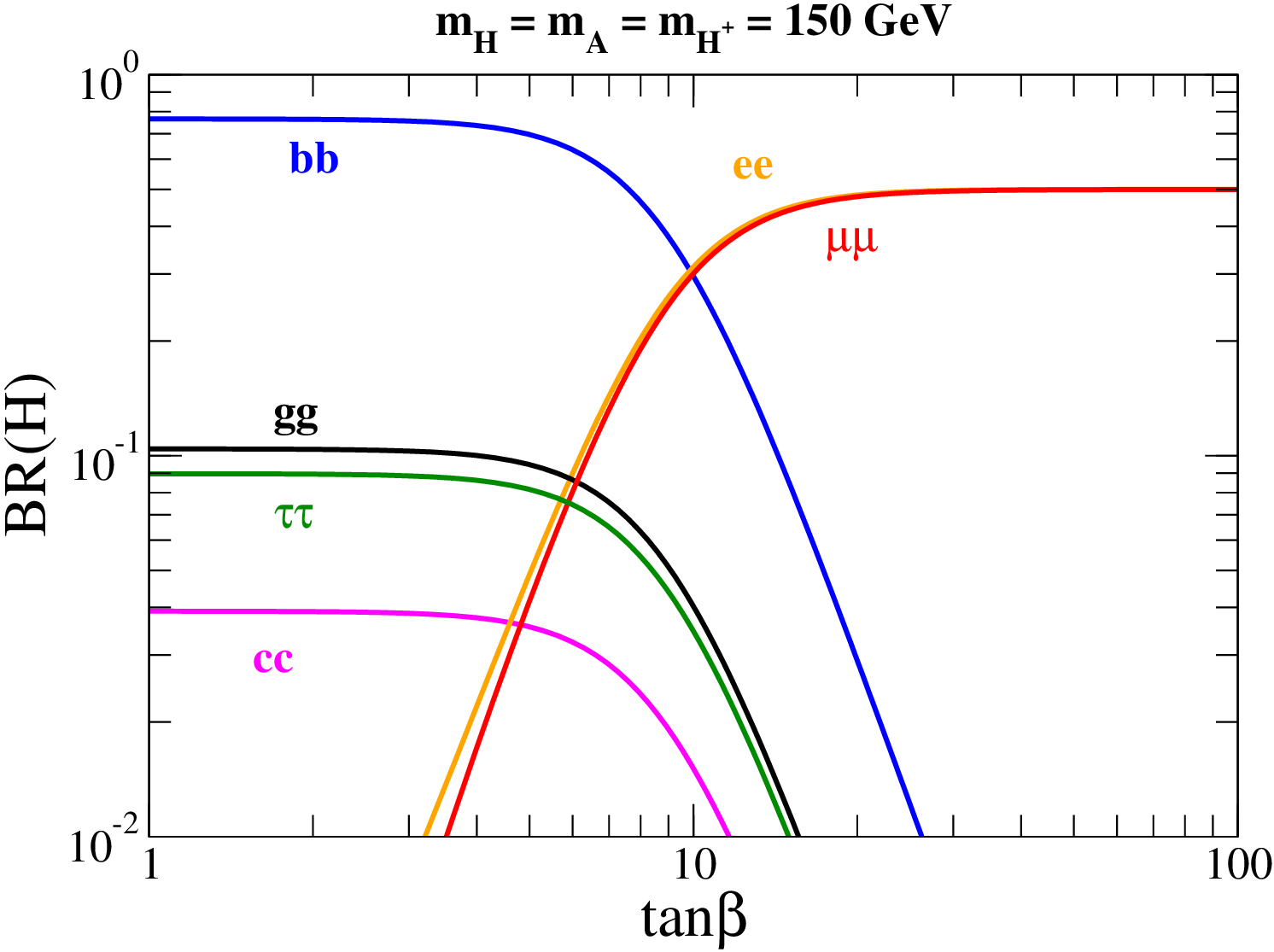}
\hspace{5mm}
 \includegraphics[width=70mm]{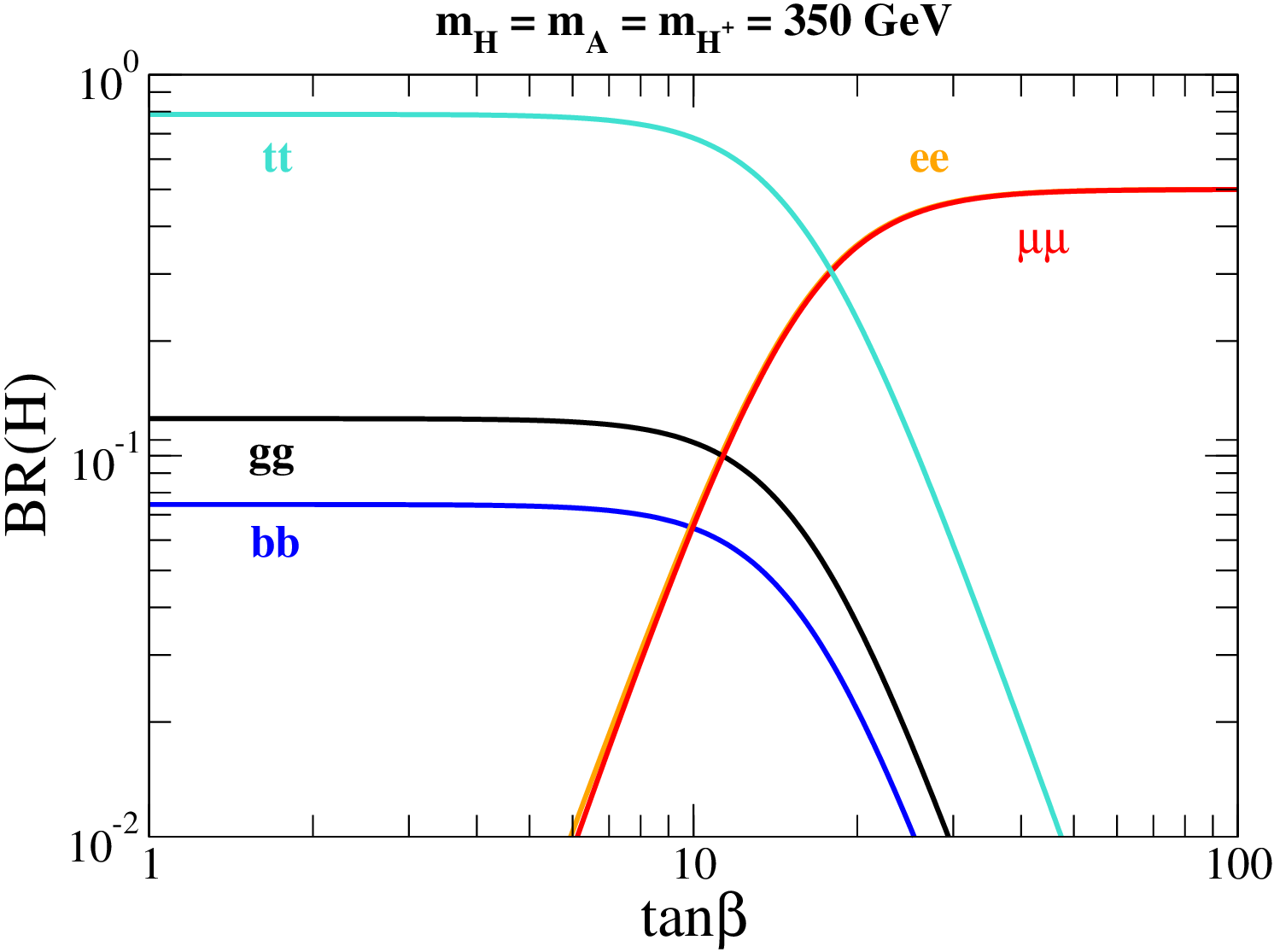}
 \caption{Decay branching ratio of $H$ as a function of $\tan\beta$ in the case with $\sin(\beta-\alpha)=1$. 
In the left and right panel, the mass of $H$ is taken to be 150 GeV and 350 GeV, respectively.}
   \label{decay1}
\end{center}
\end{figure}

\begin{figure}[t]
\begin{center}
 \includegraphics[width=70mm]{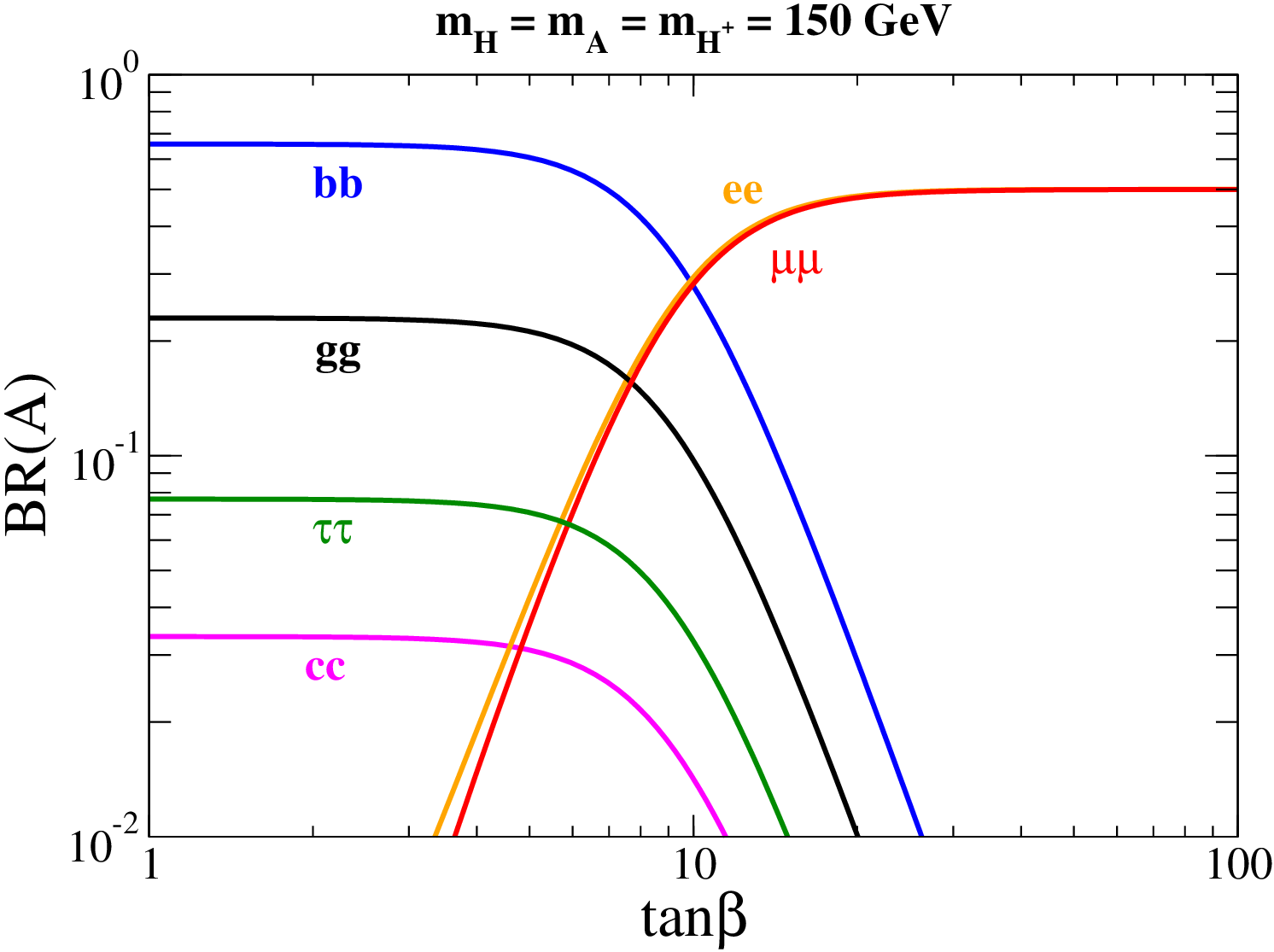}
\hspace{5mm}
 \includegraphics[width=70mm]{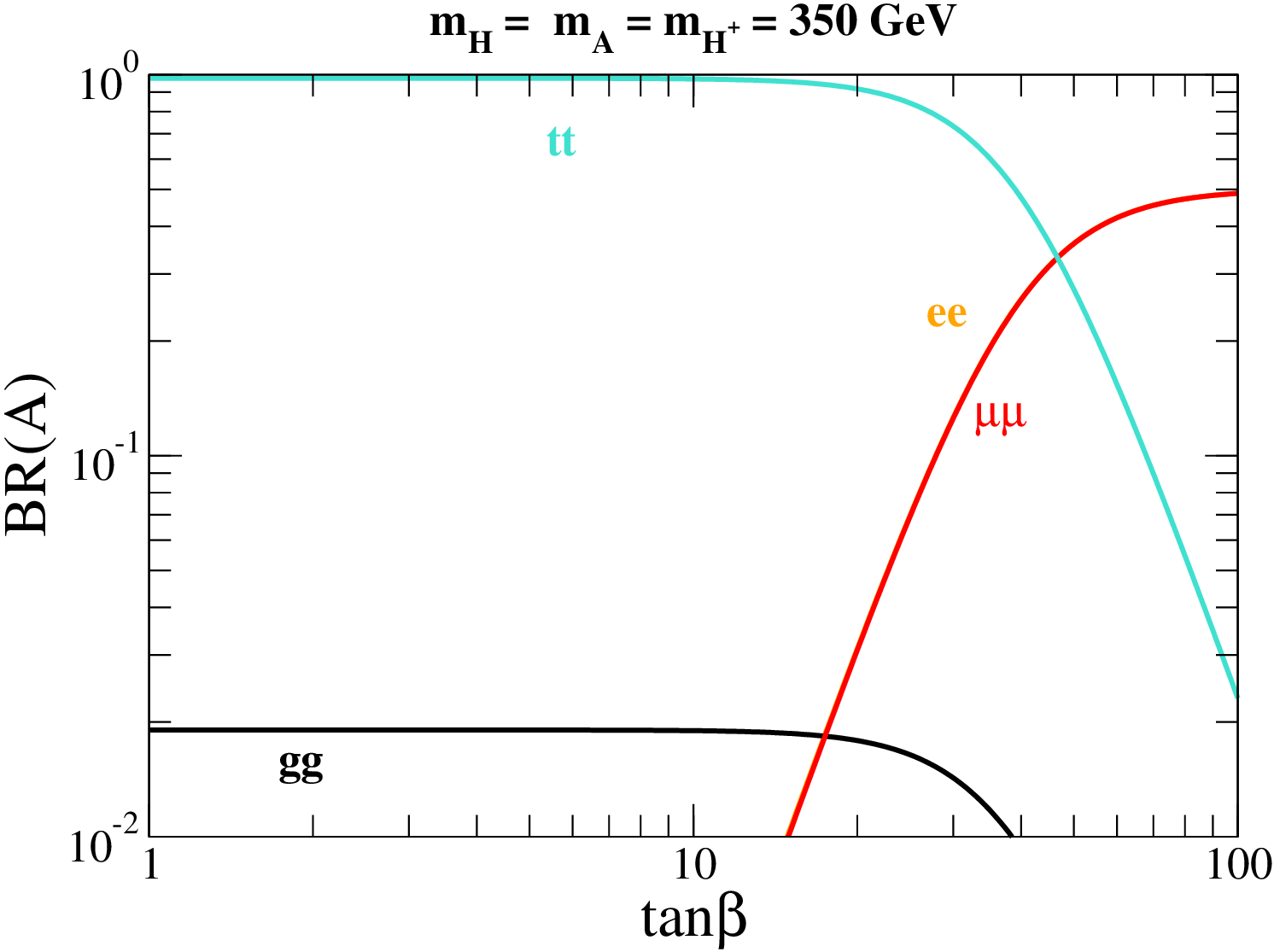}
 \caption{Decay branching ratio of $A$ as a function of $\tan\beta$ in the case with $\sin(\beta-\alpha)=1$. 
In the left and right panel, the mass of $A$ is taken to be 150 GeV and 350 GeV, respectively. }
   \label{decay2}
\end{center}
\end{figure}

\begin{figure}[h]
\begin{center}
 \includegraphics[width=70mm]{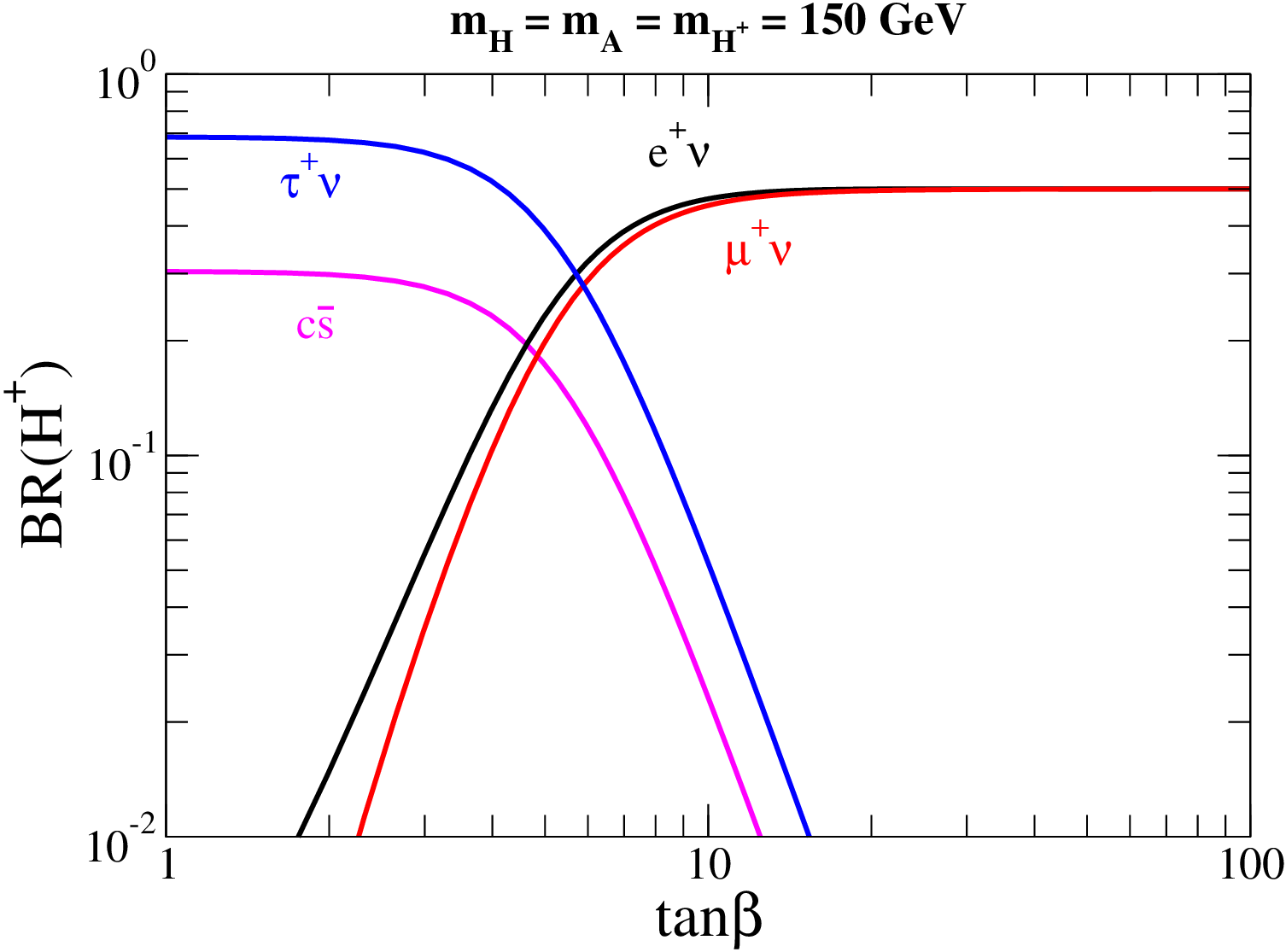}
\hspace{5mm}
 \includegraphics[width=70mm]{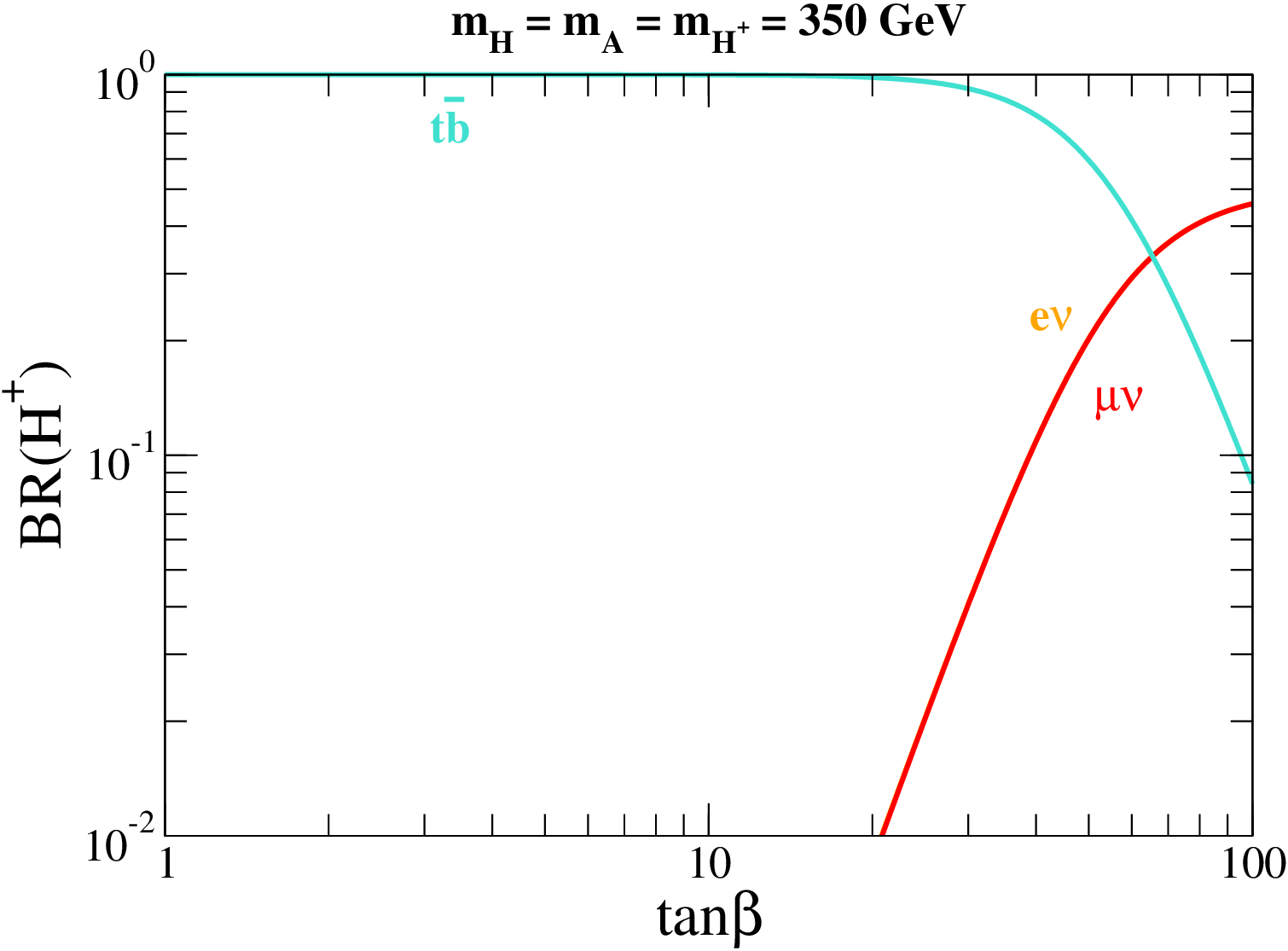}
 \caption{Decay branching ratio of $H^+$ as a function of $\tan\beta$ in the case with $\sin(\beta-\alpha)=1$. 
In the left and right panel, the mass of $H$ is taken to be 150 GeV and 350 GeV, respectively. }
   \label{decay3}
\end{center}
\end{figure}

We first evaluate the decay branching ratios of $H$, $A$ and $H^\pm$ in the Type-I $\mathbb{S}_3$ model.  
In the following calculation, the running quark masses are taken to be 
$\bar{m}_b=3.0$ GeV, $\bar{m}_c=0.677$ GeV and $\bar{m}_s=0.0934$ GeV. 
The top quark mass is set to be 173.1 GeV. The strong coupling constant $\alpha_s$ is fixed by 0.118. 
In Fig.~\ref{decay1}, 
the decay branching fraction of $H$ is shown as a function of $\tan\beta$ in the case of 
$m_H=150$ GeV (left panel) and 350 GeV (right panel).  
In the small $\tan\beta$ region, the main decay modes are $b\bar{b}$ ($t\bar{t}$), while
they are replaced by $\mu^+\mu^-$ and $e^+e^-$ when $\tan\beta$ is larger than about 10 (20)
in the case of 150 GeV (350 GeV). 

In Fig.~\ref{decay2}, the decay branching fraction of $A$ is shown as a function of $\tan\beta$ in the case of 
$m_A=150$ GeV (left panel) and 350 GeV (right panel).   
The $\tan\beta$ dependence of the branching fraction is not so different from that of $H$ in the case of 150 GeV. 
On the other hand, in the case of $m_A=350$ GeV, the meeting point of two curves for $t\bar{t}$ and $e^+e^-$ or $\mu^+\mu^-$ 
is shifted into the larger $\tan\beta$ value about 50, 
because the suppression of the decay rate of $A\to t\bar{t}$ due to the  
phase space function is weaker than that of $H$. 
 
The branching fraction of $H^+$ is shown in Fig.~\ref{decay3} as a function of $\tan\beta$ in the case of 
$m_{H^+}=150$ GeV (left panel) and 350 GeV (right panel).
When $\tan\beta\lesssim 7$ ($\tan\beta>7$), the $H^+ \to \tau^+\nu$ ($H^+ \to \mu^+\nu$ and $e^+\nu$) 
decay is dominant in the case of $m_{H^+}=150$ GeV. 
When $m_{H^+}=350$ GeV, the main decay mode is changed from $t\bar{b}$ to $\mu^+\nu$ and $e^+\nu$ at $\tan\beta\simeq 65$. 

We would like to mention that measuring almost the same branching fractions of $H/A\to e^+e^-$ and $H/A\to \mu^+\mu^-$ as well as those of
$H^+\to e^+\nu$ and $H^+\to \mu^+\nu$ can be an evidence of the $\mathbb{S}_3$ symmetric nature of the model; namely, 
the electron and muon are included in the same $\mathbb{S}_3$ doublet. 

\subsection{Collider signatures}

Next, we discuss signatures of the extra Higgs bosons at the LHC. 
The main production mode of $H$ and $A$ is the gluon fusion process, especially in the small $\tan\beta$ region. 
The cross section of this mode is suppressed by the factor of $\cot^2\beta$, so that it does not use in the large $\tan\beta$ region.  
On the other hand, the cross section for the pair production processes $pp\to HA$, $H^\pm H$ and $H^\pm A$ do not depend on $\tan\beta$, 
so that they can be useful even in the large $\tan\beta$ region. 
We note that the vector boson fusion processes for $H$ and $A$ are vanished at the tree level in the scenario based on $\sin(\beta-\alpha)=1$. 
Thus, we consider the signal events from the gluon fusion and the pair production processes. 

\begin{table}[t]
\centering {\fontsize{10}{12}
\begin{tabular}{c|c|c|c|c}\hline\hline
$m_{h_{\text{SM}}}$ [GeV] &  $\kappa$~\cite{ATLAS_mumu} & $\tan\beta$ ($H$) & $\tan\beta$ ($A$) & $\tan\beta$ ($H$ and $A$)\\\hline
110 & 5.1 & 5.0-16.8 & 3.3-28.1 & 3.0-33.3 \\\hline
115 & 5.7 & 5.3-16.2 & 3.3-27.4 & 3.0-32.3 \\\hline
120 & 9.2 & 6.6-12.2 & 4.0-22.1 & 3.6-26.1 \\\hline
125 & 9.8 & 6.2-12.9 & 4.0-22.8 & 3.3-27.1 \\\hline
130 & 10.8 & 6.3-13.5 & 4.0-23.4 & 3.3-27.7 \\\hline
135 & 11.0 & 5.6-15.2 & 3.6-25.8 & 3.3-30.4 \\\hline
140 & 16.8 & 6.0-13.5 & 4.0-23.4 & 3.3-28.1 \\\hline
145 & 16.9 & 5.0-16.5 & 3.6-27.7 & 3.0-32.7 \\\hline
150 & 22.1 & 4.6-17.8 & 3.3-29.7 & 3.0-35.0 \\\hline\hline
\end{tabular}
} 
\caption{$\kappa$ values and the excluded range of $\tan\beta$ with the 95\% C.L. for each mass of the SM Higgs boson. }
\label{tanb_const}
\end{table}

From the gluon fusion process, the opposite-sign dimuon or dielectron signal can be considered as
\begin{align}
gg\to H/A\to \ell^+\ell^-, \label{os_mu}
\end{align}
where $\ell^\pm$ are $e^\pm$ or $\mu^\pm$. 
The cross section for this process for $\ell^\pm =\mu^\pm$ is constrained by using the analysis of 
the search for the SM Higgs boson in the dimuon decay which 
has been performed from the ATLAS data~\cite{ATLAS_mumu} with the collision energy to be 8 TeV and the integrated luminosity to be 
20.7 fb$^{-1}$. 
The current 95\% C.L. upper limit for the cross section $\sigma(pp\to h\to \mu^+\mu^-)_{\text{95\%~C.L.}}$ is given by 
$\sigma(pp\to h\to \mu^+\mu^-)_{\text{SM}}\times \kappa$, where 
$\sigma(pp\to h\to \mu^+\mu^-)_{\text{SM}}$ is the SM prediction of the cross section of the $pp\to h\to \mu^+ \mu^-$ process. 
The $\kappa$ values are listed for each mass of the SM Higgs boson $m_{h_{\text{SM}}}$ in Table~\ref{tanb_const}. 
In the $\mathbb{S}_3$ model, this cross section with the $H$ and $A$ mediations can be calculated by 
\begin{align}
\sigma(gg\to H/A\to \mu^+\mu^-)=\sigma(gg\to h)_{\text{SM}}\frac{\Gamma(gg\to H/A)}{\Gamma(gg\to h)_{\text{SM}}}
\times \text{BR}(H/A\to \mu^+\mu^-), \label{gf_HA}
\end{align} 
where $\sigma(gg\to h)_{\text{SM}}$ is the gluon fusion cross section for the SM Higgs boson , 
$\Gamma(gg\to h)_{\text{SM}}$ $[\Gamma(gg\to H/A)]$ is the decay rate of the SM Higgs boson [$H/A$] into two gluons, and 
$\text{BR}(H/A\to \mu^+\mu^-)$ is the branching fraction of the dimuon decay of $H/A$. 
In order to obtain the cross section from Eq.~(\ref{gf_HA}), 
the masses of  $H$ and $A$ are taken to be the same as that of the SM Higgs boson. 
We use the value of $\sigma(gg\to h)_{\text{SM}}$ from Ref.~\cite{gfusion8} with the 8 TeV energy. 
We then obtain the excluded ranges of $\tan\beta$ for the given values of $m_H$ and $m_A$ by requiring 
\begin{align}
\sigma(pp\to h\to \mu^+\mu^-)_{\text{95\%~C.L.}}>\sigma(gg\to H/A\to \mu^+\mu^-). \label{bound1}
\end{align}

In Table~\ref{tanb_const}, excluded ranges of $\tan\beta$ with the 95\% C.L. are listed by using Eq.~(\ref{bound1}) for each $\kappa$ value.  
In this table, the values written 
in the third, fourth and last columns respectively show the excluded range of $\tan\beta$ only by taking into account 
the $H$, $A$ contribution and both $H$ and $A$ contributions with $m_H=m_A$ to the dimuon process. 
We find that the region of $3\lesssim \tan\beta\lesssim 30$ is excluded with the 95\% C.L. in the mass range from 110 GeV to 150 GeV 
in the case of $m_H=m_A$. 

\begin{table}[t]
\centering {\fontsize{10}{12}
\begin{tabular}{c|ccccccccccc}\hline\hline
$m_A$ [GeV]  & 100  & 120  & 140  & 160  & 180  & 200  & 250  & 300  & 400  & 500  \\\hline
$HA$ [fb]  & 81.7 & 39.4 & 21.0 & 12.1 & 7.46 & 4.75 & 1.76 & 0.74 & 0.17 & 0.05 \\\cline{2-11}
& (231) & (118) & (66.4) & (40.6) & (26.1) & (17.5) & (7.43) & (3.62) & (1.10) & (0.41) \\\hline
$H^+H$ [fb]  & 95.8 & 47.6& 26.4  & 15.6 & 9.76 & 6.31 & 2.45 & 1.08 & 0.26 & 0.07 \\\cline{2-11}
  & (253) & (133) & (76.5) & (47.7) & (31.2) & (21.3) & (9.28) & (4.66) & (1.48) & (0.57) \\\hline
$H^-H$ [fb]  & 49.3 & 23.4 & 12.3 & 6.97 & 4.19 & 2.63 & 0.94 & 0.38 & 0.08& 0.02  \\\cline{2-11}
 & (152) & (76.4) & (42.8) & (25.8) & (16.4) & (10.9) & (4.49) & (2.12) & (0.61) & (0.22)  \\\hline\hline
\end{tabular}
}
\caption{Cross sections for the $HA$, $H^+H$ and $H^-H$ productions for each fixed value of $m_A$ with  
the collision energy to be 7 TeV (14 TeV). 
The masses of $H$ and $H^\pm$ are taken to be the same as $m_A$. 
The $H^\pm A$ production cross sections are the same as those of $H^\pm H$. }
\label{cs1}
\end{table}

Apart from the gluon fusion process, we discuss the pair production processes. 
In Table~\ref{cs1}, the cross sections for the pair productions are listed with the collision energy to be 7 TeV and 14 TeV in 
the case of $m_H=m_A=m_{H^+}$. 
From these processes, we can obtain the same-sign dilepton events as follows
\begin{align}
pp\to HA \to \ell^+\ell^-\ell^+\ell^- ,\quad pp\to H^\pm H/H^\pm A \to  \ell^\pm\nu\ell^+\ell^-. \label{ss_mu}
\end{align}
There are three (four) possible final states; $i.e.,$ $e^+e^-e^+e^-$, $\mu^+\mu^-\mu^+\mu^-$ and $e^+e^-\mu^+\mu^-$ 
($e^\pm\nu e^+e^-$, $\mu^\pm\nu\mu^+\mu^-$, $\mu^\pm\nu e^+ e^-$ and $e^\pm\nu \mu^+ \mu^-$) for 
the $HA$ ($H^\pm H/H^\pm A$) production mode. 
The same-sign dilepton event search has been reported by the ATLAS Collaboration 
with the collision energy to be 7 TeV and the integrated luminosity to be 4.7 fb$^{-1}$ in~\cite{ATLAS_ss}. 
The strongest constraint can be obtained from the $\mu^+\mu^+$ event whose 95\% C.L. upper limit for the cross section is 
given by 15.2 fb. 
According to Ref.~\cite{ATLAS_ss}, we impose the following kinematic cuts which are used to obtain the above upper bound as 
\begin{align}
&|\eta^{\ell}|<2.5,\quad p_T^{\ell}>20\text{ GeV}, \label{bc}\\
&M_{\ell\ell}>15~\text{GeV}, \label{inv}
\end{align}
where $\eta^{\ell}$, $p_T^{\ell}$ and $M_{\ell\ell}$ are the pseudorapidity, the transverse momentum for a charged-lepton and the
invariant mass for a dilepton system, respectively. 
In order to compare the upper limit for the cross section of the $\mu^+\mu^+$ channel, 
the above cuts should be imposed for $\ell=\mu^+$. 
The signal cross sections are calculated by using {\tt CalcHEP}~\cite{CalcHEP} and {\tt Cteq6l} for the parton distribution function (PDF). 

\begin{table}[t]
\begin{center}
\centering {\fontsize{10}{12}
\begin{tabular}{c|ccccccccccc}\hline\hline
$m_A$ [GeV]  & 100  & 110 & 120  & 130 & 140  & 150 & 160  &170 & 180  &190& 200    \\\hline
$\mu^+\mu^-\mu^+\mu^-$ [fb]  & 59.5 & 42.8 & 31.4 & 23.4 & 17.7 & 13.6 & 10.1 & 8.35 & 6.68 & 5.37&4.32\\\hline
$\mu^+\mu^-\mu^+\nu$ [fb]  & 67.8 & 49.9 & 37.3 & 28.5 & 21.8 & 17.1 & 13.4 & 10.8 & 9.05 & 7.06 & 5.74\\\hline
$\mu^+\mu^+X$ [fb]  & 195 & 143 &106 &80.3 &61.4 &47.9 &37.3 &29.9 &24.0 &19.5 &15.8\\\hline\hline
\end{tabular}
\caption{Cross sections for the $pp\to HA\to \mu^+\mu^-\mu^+\mu^-$ and $pp\to H^+H\to \mu^+\mu^-\mu^+\nu$ processes 
with the collision energy to be 7 TeV
after taking the kinematic cuts given in Eqs.~(\ref{bc}) and (\ref{inv}) for $\ell=\mu^+$. 
The total cross section of the $\mu^+\mu^+X$ final states are also shown in the last row. 
The masses of $H$ and $H^\pm$ are taken to be the same as $m_A$. 
The branching fractions of $H/A\to \mu^+\mu^-$ and $H^+\to \mu^+\nu$ are taken to be 100\%. }
\label{cs2}
}
\end{center}
\end{table}

In Table~\ref{cs2}, the cross sections for the $pp\to HA\to \mu^+\mu^-\mu^+\mu^-$ and $pp\to H^+H\to \mu^+\mu^-\mu^+\nu$ are listed 
after taking the cuts given in Eqs.~(\ref{bc}) and (\ref{inv}) for $\ell=\mu^+$ for
each fixed value of $m_A$ with the collision energy to be 7 TeV. 
We take $m_H$ and $m_{H^+}$ to be the same as $m_A$. 
The total cross section of $\mu^+\mu^+X$ final states are also shown, which is the sum of the contributions from $HA$, $H^+H$ and $H^+A$ 
productions. 
The values of the cross sections in this table are displayed by assuming 100\% branching fractions of $H/A\to \mu^+\mu^-$ and 
$H^+\to \mu^+\nu$, so that the actual cross sections are obtained by multiplying the branching fractions of the above modes.

\begin{figure}[t]
\begin{center}
 \includegraphics[width=80mm]{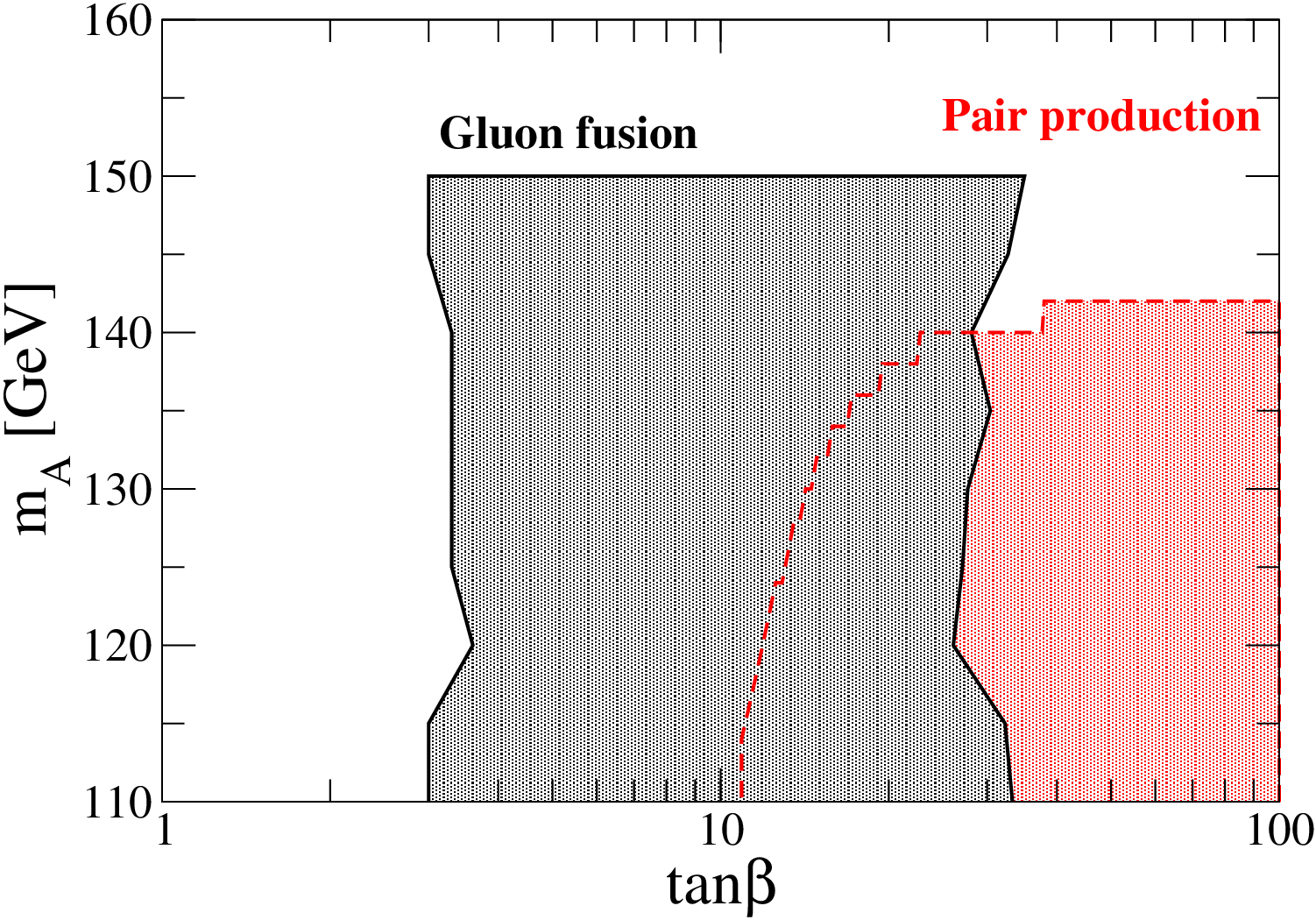}
 \caption{Excluded regions with 95\% C.L. on the $\tan\beta$-$m_A$ plane from the gluon fusion process and 
the same-sign dimuon processes at the LHC.}
\label{bound_gfusion}
\end{center}
\end{figure}

In Fig.~\ref{bound_gfusion}, the excluded regions are shown on the $\tan\beta$-$m_A$ plane in the case of $m_H=m_A=m_{H^+}$.  
The black and red shaded regions are respectively excluded with the 95\% C.L. from the opposite-sign dimuon signal from the gluon fusion process 
and the same-sign dimuon signal from the pair production processes. 
We note that the region with $\tan\beta>100$ is not so changed from that with $\tan\beta\gtrsim 30$ in this plot, because 
the branching fraction of $H/A\to \mu^+\mu^-$ and $H^+\to \mu^+\nu$ are reached to be the maximal value, $i.e.$, 50\%.  
Thus, when $m_A$ is smaller than about 140 GeV, $\tan\beta\gtrsim 3$ is excluded with the 95\% C.L. from the both constraints.  

\begin{table}[t]
\centering {\fontsize{10}{12}
\begin{tabular}{c|ccccccccccc}\hline\hline
$m_A$ [GeV]  & 100  & 120  & 140  & 160  & 180  & 200  & 250  & 300  & 400  & 500     \\\hline
$\mu^+\mu^-e^+e^-$ [fb]  & 205&123&77.8&51.4&35.3&24.8&11.5&5.858&1.88&0.72   \\\hline\hline
\end{tabular}
}
\caption{Cross sections for the  $pp\to HA\to \mu^+\mu^-e^+e^-$ process
after taking the basic kinematic cuts given in Eq.~(\ref{bc}) with the collision energy to be 14 TeV.}
\label{cs3}
\end{table}

Finally, we discuss the discovery potential of $H$ and $A$ with the collision energy to be 14 TeV. 
We focus on the pair production process, especially for the $pp\to HA\to e^+e^-\mu^+\mu^-$ event, because we can clearly see the 
electron and muon specific nature of $H$ and $A$. 
To estimate the background cross section, we use the {\tt MadGraph5}~\cite{MG5} and {\tt Cteq6l} for the PDF. 
After we impose the basic kinematic cuts as given in Eq.~(\ref{bc}) in which $\ell$ is all the charged-leptons in the final state, 
we obtain the background cross section to be about 8.1~fb. 
The signal cross section is calculated by using {\tt CalcHEP} and {\tt Cteq6l}. 
In Table~\ref{cs3}, the cross section for the $pp\to HA\to \mu^+\mu^-e^+e^-$ process after taking the kinematic cut is shown for each fixed value 
of $m_A$. 
We here introduce the signal significance $\mathcal{S}$ defined as 
\begin{align}
\mathcal{S} = \frac{ N_{\text{sig}} }{ \sqrt{N_{\text{sig}}+N_{\text{bg}}} },
\end{align}
where $N_{\text{sig}}$ and $N_{\text{bg}}$ denote the event number of the signal and background processes, respectively.

\begin{figure}[t]
\begin{center}
 \includegraphics[width=80mm]{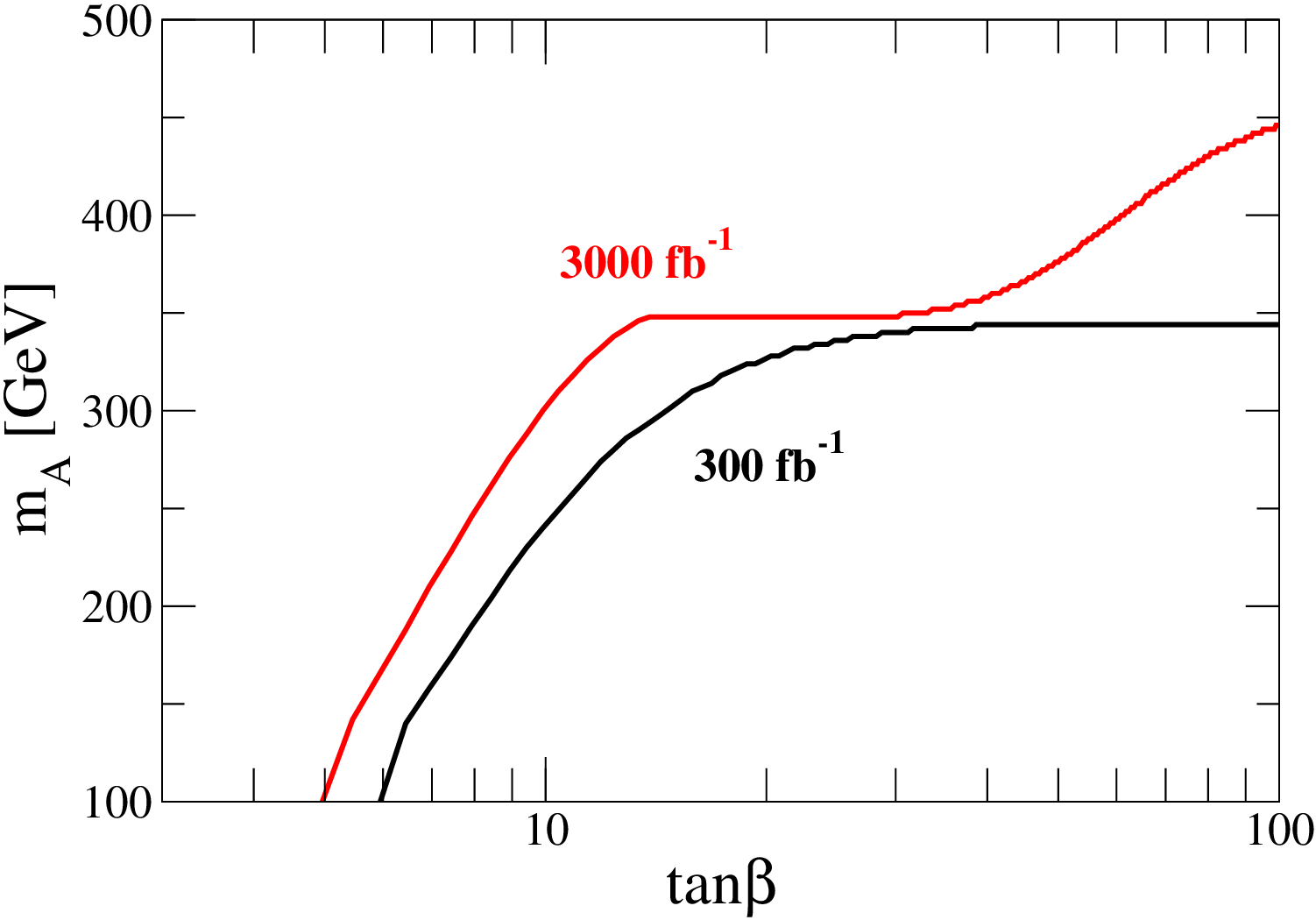}
\end{center}
 \caption{The 5$\sigma$ discovery potential at the LHC with the collision energy to be 14 TeV. 
The black and red contours respectively show the parameter region giving $\mathcal{S}=5$ by assuming the integrated luminosity to be 300 fb$^{-1}$ and 3000 fb$^{-1}$.}
\label{discovery}
\end{figure}

In Fig.~\ref{discovery}, we show the discovery potential of the $e^+e^-\mu^+\mu^-$ signal from the $pp\to HA$ production. 
The signal significance $\mathcal{S}$ is larger than 5 in the regions inside the black and red curves, where 
the integrated luminosity is assumed to be 300 fb$^{-1}$ and 3000 fb$^{-1}$. 
Because the top quark pair decay of $H$ and $A$ opens, the discovery reach is saturated at about 350 GeV. 
We find that $H$ and $A$ with their masses up to 350 GeV can be discovered by 5$\sigma$ in the case of $\tan\beta\gtrsim 30$ with 300 fb$^{-1}$. 
In the 3000 fb$^{-1}$ luminosity, 
the discovery reach can be above 350 GeV when $\tan\beta\gtrsim 30$.

\section{Summary and Conclusion}

We have studied the THDMs in the framework based on the $\mathbb{S}_3$ flavor symmetry. 
Assigning the first and second generation lepton fields (two Higgs doublet fields) to be the doublet (singlet) under $\mathbb{S}_3$, 
the mass matrix for the charged-leptons is obtained to be the diagonal form in the weak eigenbasis. 
The quark masses and mixings are explained as the same way in the SM by assuming the $\mathbb{S}_3$ charge for quarks to be the singlet. 
The $\mathbb{S}_3$ charge assignment to the Higgs doublet fields in our model, 
which is different from the previous studies for $\mathbb{S}_3$ models where the Higgs fields 
are usually taken to be the $\mathbb{S}_3$ doublet, is suitable to explain   
the SM-like Higgs boson with the mass of 126 GeV discovered at the LHC. 

The tree level FCNC appearing in the general THDMs is forbidden by the $\mathbb{S}_3$ symmetry in our model set up in which
four types of the Yukawa interaction are allowed depending on the $\mathbb{S}_3$ charge assignments for 
fermions named as Type-I, Type-II, Type-X and Type-Y $\mathbb{S}_3$ models. 
We have found that the extra Higgs bosons $H$, $A$ and $H^\pm$ 
can be electron and muon specific in the Type-I $\mathbb{S}_3$ model in the  
large $\tan\beta$ regions. 
Namely, 
the decay modes of $H/A\to \mu\mu$, $H/A\to ee$ and $H^\pm\to \mu^\pm\nu/e^\pm \nu$ are dominant, 
and the branching fraction for the muon final state is almost the same as that for the electron final state.  
This property does not appear in any other THDMs without the tree level FCNC such as 
a ${\mathbb Z}_2$ symmetric version of the THDMs. 
Therefore, measuring signatures of the $\mu/e$ specific extra Higgs bosons can be a direct probe of our model. 

We have explored excluded regions on the $\tan\beta$-$m_A$ plane has been evaluated as shown in Fig.~\ref{bound_gfusion} 
by using the Higgs boson search data of the dimuon decay mode data and the same-sign dimuon event. 
We also have estimated the 5$\sigma$ discovery potential of the $pp\to HA\to e^+e^-\mu^+\mu^-$ signal assuming the center of mass energy 
to be 14 TeV and the integrated luminosity to be 300 fb$^{-1}$ and 3000 fb$^{-1}$.

\section*{Acknowledgments}
%\vspace{0.5cm}
H.O. thanks to Prof. Eung-Jin Chun for fruitful discussion.
Y.K. thanks Korea Institute for Advanced Study for the travel support and local hospitality
during some parts of this work.
K.Y. was supported in part by the National Science Council of R.O.C. under Grant No. NSC-101-2811-M-008-014.

\end{document}